\documentclass[]{aa}
\usepackage{graphicx}
\usepackage{txfonts}
\begin{document}
\title{Recovery of the old nova V840 Ophiuchi: A carbon rich system
\thanks{Based on observations collected at the European Southern Observatory, 
        La Silla, Chile}}
\titlerunning{Recovery of the carbon--rich nova V840 Ophiuchi}
\author{L. Schmidtobreick \inst{1}
        \and
        C. Tappert \inst{2}
        \and
        A. Bianchini \inst{3}
        \and
        R.\,E. Mennickent \inst{2}}
\offprints{Linda Schmidtobreick, \email{lschmidt@eso.org}}
\institute{European Southern Observatory, Casilla 19001, Santiago 19, Chile.
           \and
           Grupo de Astronom\'{\i}a, Universidad de Concepci\'on,
           Casilla 160--C, Concepci\'on, Chile
           \and
           Dipartimento di Astronomia, Universit\`a di Padova,
           Vicolo dell'Osservatorio 2, I-35122, Padova, Italy}
\date{Received xxx xxx, xxx; accepted xxx xxx, xxx}
\abstract{We present optical spectroscopy and multi colour photometry
of the old nova V840 Oph. We rediscovered the nova based on its position
in the colour--colour diagrams. It stands out as a very blue 
object with an additional red component. 
We present the first optical spectroscopy of this candidate and confirm
its nova character.
Furthermore, V840\,Oph has been found as one of very few cataclysmic variables
showing C\,IV emission at $\lambda = 580/1$\,nm. From the analysis of the 
carbon lines it seems probable that V840\,Oph contains actually a 
carbon--rich secondary star. So far, only the nova--like QU\,Car has been known
to have such a companion.
We furthermore find spectroscopic evidence that V840\,Oph has a hot, dense accretion
disc or stream and is probably a magnetic system.
   }
\maketitle

\section{Introduction}
V840\,Oph, a classical Galactic nova, has first been visible as a 
star of 6.5\,mag on Harvard
photographs taken in May 1917 (Bailey \cite{bail20}). 
It has been a fast nova with
two very distinct secondary maxima (Shapley \cite{shap21}),
and has been classified by Duerbeck (\cite{duer81})
as Bb (decline with major fluctuations). He also gave the value
for $t_3$ = 36\,d which places V840\,Oph among the fast novae.
Three stars are close to the position of the nova and possible candidates
for the nova remnant (Duerbeck \cite{duer87}).

Several surveys have attempted to identify V840 Oph, but without much success.
No detection has been accomplished in the 2mass second incremental data release 
(Hoard et al. \cite{hoar+02}).
Munari \& Zwitter (\cite{muna+98}) performed spectrophotometric measurements
for 20 faint cataclysmic variables but V840 Oph has been too faint to
be included.
Harrison \& Gehrz (\cite{harr+94}) have observed several novae in four IR bands
with IRAS and obtained a detection of V840 Oph at 25$\mu$m.
However, due to the large FWHM of the IRAS signal (the uncertainty of the
coordinates is 0.\arcmin 5 for 25$\mu$m) source confusion is very likely, and
no unambiguous identification can be achieved in this crowded field.

In the course of a long-term project investigating novae with large outburst
amplitudes, we have 
performed multi wavelength photometry of the stars in the field 
of V840 Oph and in particular of the three candidates for the nova. 
The aim was
to recover the nova via its colour characteristics and to confirm it 
spectroscopically. In a further investigation of the system, we also
analyse its spectroscopic peculiarities.
\\
\section{Observation and data reduction}
\begin{table}[b]
\caption{\label{obstab} All observations obtained for this research are listed with
their date, the instrument/telescope combination, the used filter or grism and 
slit width, and the exposure time.}
\begin{tabular}{c c c c}
\hline
\hline
 \noalign{\smallskip}
Date & Instrument & Filter or Grism/Slit & Exp-time [s] \\
 \noalign{\smallskip}
\hline
 \noalign{\smallskip}
2001-07-15 & DFOSC/1.54D & U & 900  \\
2001-07-15 & DFOSC/1.54D & B & 900  \\
2001-07-15 & DFOSC/1.54D & V & 600  \\
2001-07-15 & DFOSC/1.54D & R & 900  \\
2001-07-15 & DFOSC/1.54D & I & 900  \\
2002-04-20 & EFOSC/3.6   & G9/1.5" & 2$\times$1200 \\
2003-02-28 & EFOSC/3.6   & G4/1.5" & 3$\times$900 \\
 \noalign{\smallskip}
\hline
\end{tabular}
\end{table}

\begin{figure}
\resizebox{4.8cm}{!}{\includegraphics{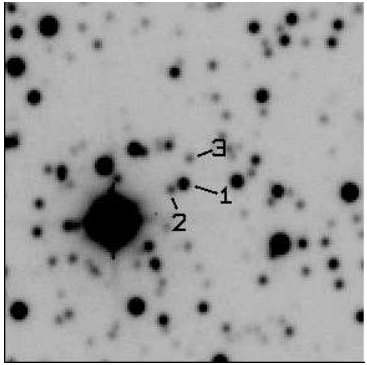}}
\caption{\label{finding} The B--band finding chart shows the 
$1 \times 1$\,arcmin 
surroundings of V840\,Oph; north is up, east is left. 
The three possible candidates for the old nova are indicated 
by the numbers that have been assigned by Duerbeck (\cite{duer87}).}
\end{figure}

\begin{figure*}
\rotatebox{-90}{\resizebox{!}{18.1cm}{\includegraphics{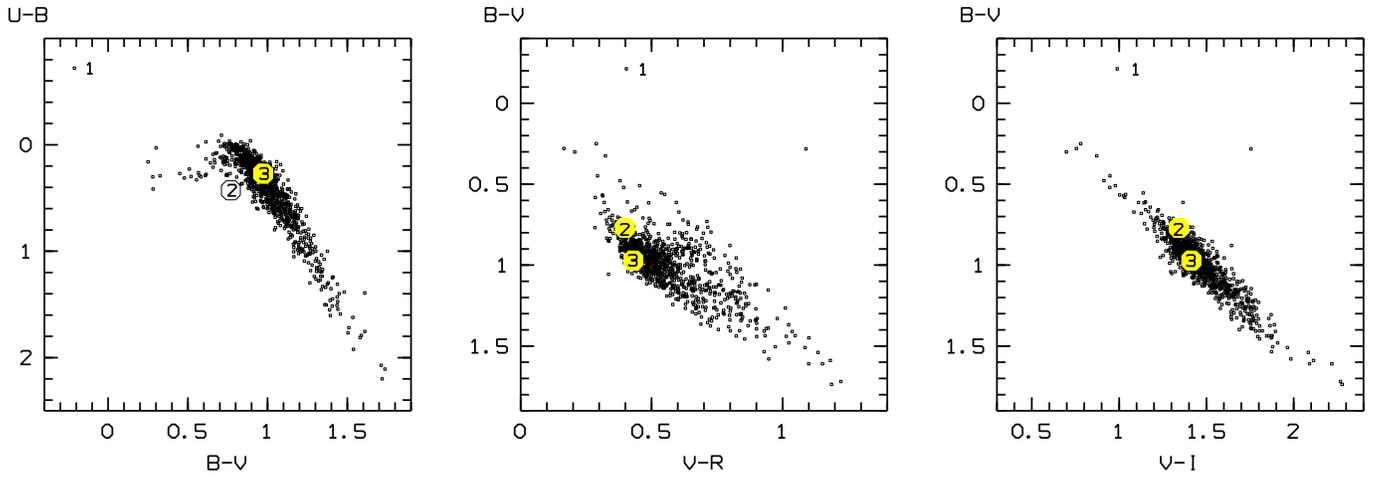}}}
\caption{\label{2col} Colour--colour diagrams are plotted for the three nova 
candidates and the surrounding field stars. While candidates 2 and 3 lie close
to the main sequence (defined by the majority of field stars), the best 
candidate (No 1)
clearly stands out in all three diagrams.}
\end{figure*}

The photometric data were obtained in one night on 2001-07-15 at the
1.54 m Danish telescope at ESO, La Silla, using DFOSC. Spectroscopic 
observations were performed on 2002-04-20 and on 2003-02-28 using EFOSC2 at the
3.6 m telescope at La Silla. In Table \ref{obstab} the observational
parameters are summarised. 

The standard reduction of the photometric data
including BIAS subtraction and flatfielding have been performed using IRAF. 
For the stellar photometry, the standalone packages of DAOPHOT and ALLFRAME
have been used to perform aperture and PSF photometry of all stars in the 
observed field. Only those stars with photometric uncertainties below 0.03 mag
have been selected for further analysis.
Photometric calibration has been obtained with MIDAS using
several standard stars selected from Landolt (\cite{land92}).

The reduction of the spectroscopic data has been done with IRAF only.
The BIAS has been subtracted and the data have been divided by a flat field,
which was normalised by fitting Chebyshev functions of high order. 
The spectra have been optimally extracted (Horne \cite{horn86}).
Wavelength calibration yielded a final resolution of 0.92\,nm\ FWHM
for the 2002 data and 1.63\,nm\ FWHM for the 2003 data.
Flux calibration was performed only for the 2003 spectrum, using the
spectrophotometric standard LTT\,6248 which has been observed directly 
after the object with an airmass difference of 0.1. The photometric error has
been derived by comparing the three individual spectra and the photometry
of the acquisition image and is estimated as 
$\rm 0.2 \cdot 10^{-18}W m^{-2} nm^{-1}$.

For the further analysis of both the photometric and spectroscopic data,
the MIDAS package and self--written routines have been used.

\section{The method of recovery}
Cataclysmic variables are close, interacting binary systems, 
comprising a white dwarf accreting mass
from a Roche--lobe--filling late--type
star.  Hence there are at least three different physical components 
that provide the light observed from a cataclysmic variable.
Due to its size and high temperature, the accretion disc or stream is 
generally the strongest in the optical range. While the white dwarf contribution
affects mainly the UV range of the spectrum, the secondary late--type star
contributes on the red and infrared side. 

The compound of these different physical components results therefore in very 
characteristic colour terms. Cataclysmic variables appear 
as generally very blue objects with a shift towards the red
at longer wavelengths, depending on the strength of the
secondary in comparison to the accretion disc or stream.
In a colour--colour diagram, 
this quality places them on the blue side but slightly above the main sequence.

Performing multi colour photometry on all stars within the field of a
suspected nova should therefore show the following picture:
The majority of the stars are gathered along a main sequence. The latter 
is shifted
from the theoretical one according to the mean interstellar reddening in 
the field, while the scatter of the stars indicates the variation of the
extinction within the field. All candidates within the coordinate uncertainty
of the nova will be marked. Assuming that the interstellar reddening of the
nova is of the same order as the other field stars, it should be located 
on the blue side above the main sequence defined by the field stars,
and can thus be identified. 

\section{Results and discussion}
\subsection{The nova candidate}
\begin{figure*}
\centerline{
\rotatebox{-90}{\resizebox{!}{14cm}{\includegraphics{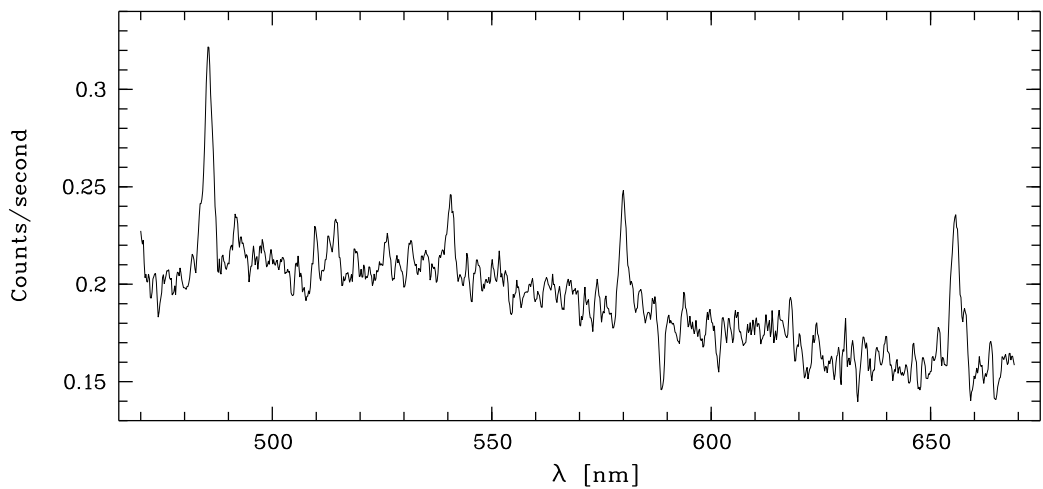}}}}
\caption{\label{v840s} Spectrum of candidate No 1. 
Its nova character is confirmed by the presence 
of typical nova emission lines. Noteworthy are the strong He\,II line at
$\lambda = 541$\,nm and the presence of the C\,IV doublet at $\lambda = 580/1$\,nm
(not resolved).}
\end{figure*}
\begin{table*}
\caption{\label{coltab} Position, magnitudes and colours for the three candidates
are given with their photometric errors.
The ID numbers are as in Fig.~\ref{finding}.}
\centerline{
\begin{tabular}{l c c c c c c c }
\hline
\hline
 \noalign{\smallskip}
ID & RA$_{2000}$ & Dec$_{2000}$ & V & U-B & B-V & V-R & V-I \\
 \noalign{\smallskip}
\hline
 \noalign{\smallskip}
1 & 16:54:43.92 & -29:37:26.8 & 19.32(3) & -0.72(2) & -0.21(3) & 0.40(3) & 0.99(4)\\
2 & 16:54:44.07 & -29:37:27.7 & 19.71(3) &  0.43(6) &  0.77(4) & 0.40(4) & 1.34(5) \\
3 & 16:54:43.85 & -29:37:22.9 & 19.95(3) &  0.27(7) &  0.97(4) & 0.98(4) & 1.41(5) \\
 \noalign{\smallskip}
\hline
\end{tabular}}
\end{table*}

Fig.~\ref{finding} shows a $1 \times 1$\,arcmin close-up of the B image 
surrounding the nova position.
The three stars that lie within the  uncertainty range of the nova coordinates
 are indicated by the same ID numbers as in Duerbeck (\cite{duer87}).
In Table \ref{coltab} the magnitudes and colours of these three objects 
are given. Fig.~\ref{2col} gives the colour--colour diagrams of all stars
in the field with the three candidates marked with their numbers. 
The field stars define the main sequence, shifted and spread due to
individual interstellar reddening effects. The mean extinction in the field
has been determined as $E_{B-V} = 0.4\pm 0.15$.
 Whereas candidates 2 and 3 lie
within the bulk of field stars,
candidate 1 clearly stands out in these diagrams.

To verify our method, we took a first spectrum of this best candidate.
The result is plotted in Fig.~\ref{v840s}. 
In spite of the low S/N, the Balmer emission lines, typical for
cataclysmic variables, are clearly visible and thus confirm the 
photometric selection. In addition, the nova character of this candidate
is supported by the presence of 
He\,II $\lambda 5412$, since this line is typically found in nova remnants 
but not in dwarf novae or nova--like variables (Warner, \cite{warn95}).
We therefore conclude that candidate 1 is indeed the old nova
V840\,Oph.

\subsection{The Carbon content}
The strong line of C\,IV at $\lambda = 580.5$\,nm is very rare in cataclysmic
variables. C\,IV 
is usually detected in the hot wind of Wolf-Rayet stars, in planetary nebulae 
or during nova outburst and is there assumed to originate from the expanding
shell. In the case of V840\,Oph, however, the principal indicator for ongoing 
outflow, emission lines with a P\,Cyg profile, is not observed. Furthermore,
too much time has passed since the nova outburst in 1917 for the nova shell
to provide a significant contribution, especially as it has been a very fast 
nova. 

Although uncommon, C\,IV has been found before in quiescence novae
like CP\,Pup (Williams \& Ferguson \cite{will+83}), V603\,Aql, V1500\,Cyg,
or HR\,Del (all in Ringwald et al. \cite{ring+96}), in the nova--like
variables PG\,1012-03 (Williams \& Ferguson \cite{will+83})
and V\,Sge (Williams \cite{will83}), and most recently in
QU\,Car (Drew et al. \cite{drew+03}).
However, in all these systems the C\,IV line $\lambda 580/1$ appears much 
weaker, the strongest one being the line in QU\,Car with an equivalent width of
0.13\,nm, which is still a factor of five smaller than our detection in 
V840\,Oph. However, we have to consider this number with caution,
since all emission lines in V840\,Oph have a two to three times higher 
equivalent width than in
QU\,Car, which can easily be explained by QU\,Car having a hotter continuum.
Even taking this into account, the strength of the C\,IV line in V840\,Oph 
is still remarkable.

Since C\,II and C\,III are common in most cataclysmic variables, 
there exist two
possible scenarios, which can explain the presence of a strong C\,IV emission 
in V840\,Oph: 

(1) The temperature of V840\,Oph or in parts of this system
is extremely high, with the result 
that most of the carbon present is actually ionised to C$^{4+}$, 
and the observed C\,IV emission line results from the 
$\rm C^{4+} \rightarrow C^{3+}$ recombination cascade.

(2) The total amount of carbon in V840\,Oph is higher than usual in 
these systems yielding corresponding high abundances for all 
carbon ions. 

The first scenario would
show in rather weak C\,II and C\,III lines and corresponding high
ionisation ratios of CIV/CII and CIV/CIII. The second scenario would
show in extremely strong C\,II and C\,III lines to keep the
ionisation ratios on a normal level in spite of the strong C\,IV line.
Naturally, a combination of both effects seems possible.

\begin{figure*}
\centerline{
\rotatebox{-90}{\resizebox{!}{13cm}{\includegraphics{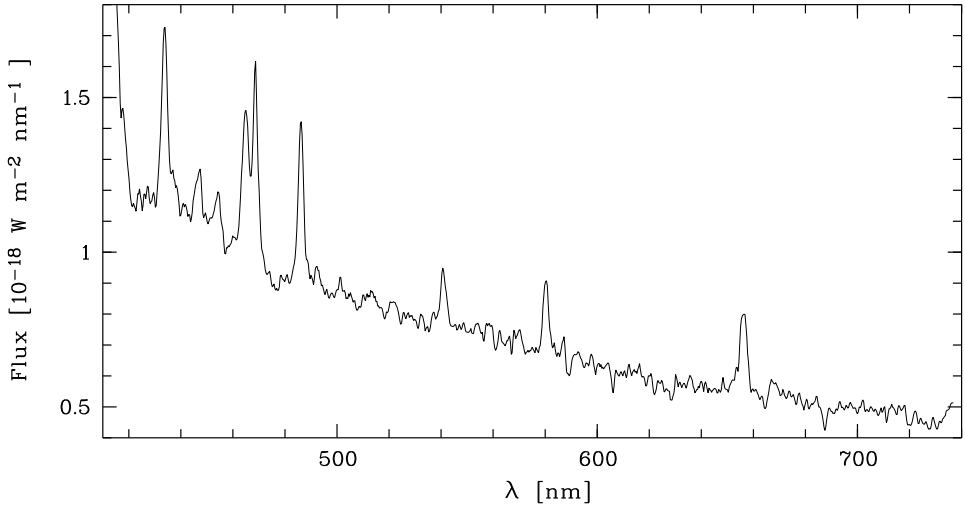}}}}
\caption{\label{aves1} Flux calibrated spectrum of V840\,Oph (candidate 1), 
obtained in February 2003. 
The spectra has not been corrected for reddening.
Apart from the high excitation lines already seen
in the 2002 spectrum, the strong emission feature between $\lambda = 460$
and 475\,nm (Bowen blend and He\,II) attracts attention. }
\end{figure*}
\begin{figure}
\rotatebox{-90}{\resizebox{!}{7.2cm}{\includegraphics{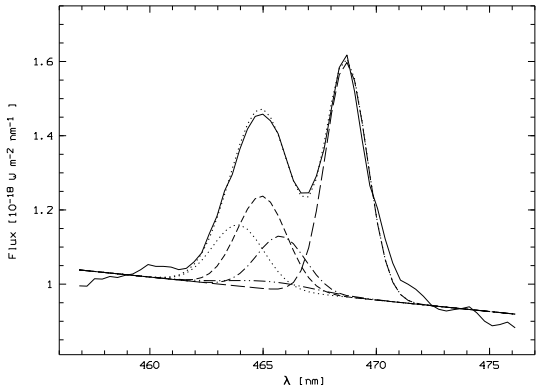}}}
\caption{\label{bowen_fit} A detailed view on the Bowen blend and
He\,II $\lambda$469. Over the original spectrum the fits to the five
ions are plotted as dotted line for N\,III, dashed line for C\,III,
dash--dot line for C\,IV, dash--dot--dot line for O\,II, and long dashes
for He\,II. The overall fit is plotted as thick dots. }
\end{figure}

To check on these possibilities,
we have taken an additional spectrum in the range 410 to 740\,nm, which 
also includes the Bowen blend (C\,III,C\,IV,N\,III) at 464\,nm 
(see Fig.~\ref{aves1}). 
Since the transmission was stable during
this night, we performed a rough flux calibration for this spectrum.
We used the photometrically determined reddening of  $E_{B-V} = 0.4$ to 
deredden 
the spectrum. However, the resulting continuum slope of $\lambda ^{-3.3}$
seems to be quite high. Since the nova as an intrinsically faint object 
might be
closer than the average of the field stars, and the scatter in the 
colour--colour diagram and hence in the derived extinction 
is quite high, we assume that $E_{B-V} = 0.4$ is an upper limit for the
nova extinction with the real value being probably lower. 
Hence, as a convention we have listed both flux values -- the ones as 
observed and the dereddened ones -- in the tables. In the further
text, however, we refer to the values as observed, if not stated 
otherwise.

We have modelled the Bowen emission feature by fitting Gaussian profiles to the
individual lines of the four ion multiplets O\,II (8 lines), N\,III 
(3 lines), C\,IV (1 line), and C\,III (3 lines) using the relative laboratory 
line strengths
and wavelengths from McClintock et al. (\cite{mccl+75}) for the components 
within each multiplet. 
A common line width has been assumed for the individual lines.
Its value 
and the four heights of the combined lines within one multiplet 
are the five parameters to fit. Since He\,II at 468.6\,nm is not completely 
separated from the blend, we have included a fit to the
He\,II part, keeping all three Gaussian parameters free for this line.
With the resulting best fit values we rebuilt the multiplets of each ion
separately to compute its FWHM and the flux contribution within. 
The results of these line--models are shown in Table \ref{bowentab} and
Fig.~\ref{bowen_fit}. The FWHM of the individual lines resulted in
$2.5 \pm 0.8$\,nm which is in good agreement with the average measured for the
other lines in the spectrum (see Table \ref{eqwtab}). 

\begin{table}
\caption{\label{bowentab} Central wavelength and line flux 
as measured and dereddened (same units)
are given of the fitted lines and blends for all ions within
the 460--472\,nm emission feature.}
\begin{tabular}{l c c c c }
\hline
\hline
\noalign{\smallskip}
Ion & $\lambda$ & F [$10^{-18}$\,W\,m$^{-2}$] & dereddened\\
 \noalign{\smallskip}
\hline
\noalign{\smallskip}
He\,II & 468.67 & 1.46 $\pm$ 0.1 & 4.03 $\pm$ 0.3\\
C\,IV  & 465.80 & 0.39 $\pm$ 0.5 & 1.60 $\pm$ 2.0\\
C\,III & 464.95 & 0.68 $\pm$ 0.2 & 2.78 $\pm$ 0.8\\
N\,III & 463.91 & 0.47 $\pm$ 0.5 & 1.92 $\pm$ 2.0\\
O\,II  & 465.43 & 0.10 $\pm$ 0.5 & 0.40 $\pm$ 2.0\\
\noalign{\smallskip}
\hline
\end{tabular}
\end{table}
\begin{figure*}
\rotatebox{-90}{\resizebox{!}{18.0cm}{\includegraphics{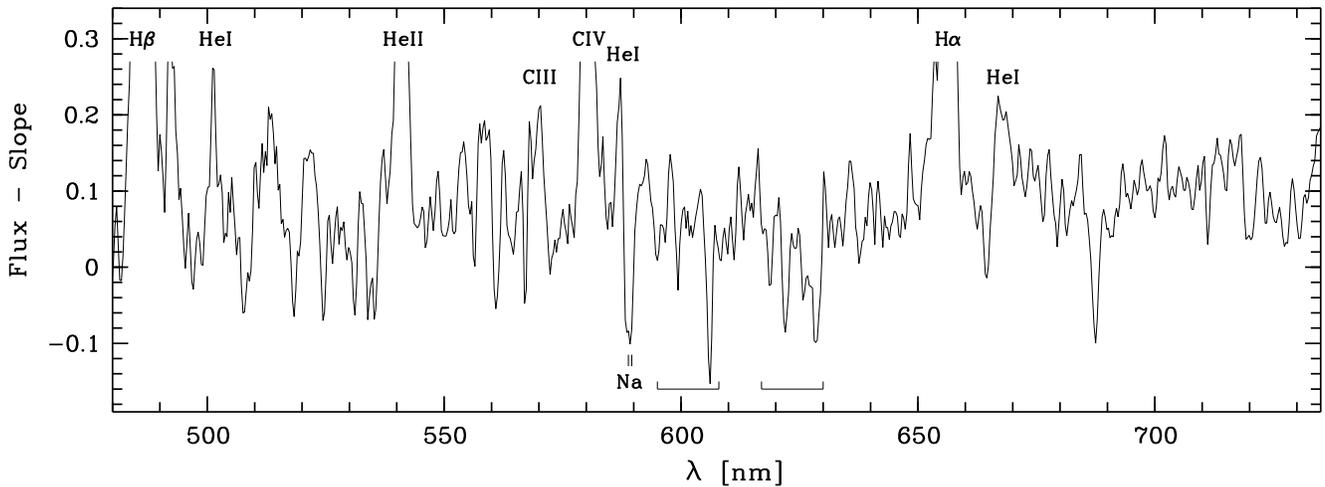}}}
\caption{\label{slope_plot}Result of subtracting a 
$\lambda^{-3.3}$ slope from the dereddened spectrum of V840\,Oph.
The identified emission lines and those
absorption features that are also present in the 2002 data are marked.  }
\end{figure*}

Other evidence for the goodness of the fit comes from the line strengths
of C\,III and C\,IV. Both ions have a relatively strong line in the
observed part of the spectrum, additionally to their contribution to the 
Bowen blend.
The ratio of effective 
recombination coefficients for C\,IV $\lambda 466$ and C\,IV $\lambda 580/1$
is 1.67 (from Kingsburgh, \cite{king+95}). 
Scaling the C\,IV $\lambda 580/1$ line accordingly, one would 
expect a flux contribution of 0.4$\cdot 10^{-18}$W\,m$^{-2}$ 
of 
C\,IV to the Bowen blend, which is in perfect agreement with our results 
(Table \ref{bowentab}).
For C\,III, the ratio of C\,III $\lambda 570$ flux and C\,III $\lambda 465$ flux
from laboratory spectra is 0.32. Scaling the fitted value of 
$F=0.68\cdot 10^{-18}$\,W\,m$^{-2}$ 
accordingly, the flux in C\,III $\lambda 570$ 
is expected around $\rm 0.22\cdot 10^{-18}\,W\,m^{-2}$, 
which is within the errors of the found 
value (see Table \ref{eqwtab}).
For the N\,III contribution, no such confirmation exists, since all other 
strong N\,III multiplets are blended with either H$\beta$, He\,II, or
H$\gamma$. At 451\,nm a small emission is seen but the noise in this part of 
the spectrum is too high for a trustworthy detection of this weak feature.
Small peaks are also found for the N\,III multiplet around 646\,nm but the
identifications are dubious due to blending with several Fe\,II multiplets in 
this region.
The low contribution of O\,II to the blend is confirmed by the absence of
features at 459.3\,nm and 470.5\,nm.

The total flux in the Bowen blend sums up to 1.64$\cdot 10^{-18}$W\,m$^{-2}$,
which is slightly higher than the flux of He\,II at 468.6\,nm. 
So far, QU\,Car has been the only known cataclysmic variable where the emission 
from the Bowen blend is comparable to the He\,II emission at 468.6\,nm.
In V840\,Oph, about two 
thirds of this flux are contributed by Carbon emission with C\,III being
about twice as strong as C\,IV. C\,III and N\,III contribute in a ratio
of about 3 to 2, while the O\,II emission is negligible. 
This appearance is similar 
to what has been found in QU\,Car (Gilliland \& Phillips \cite{gill+82})
with nitrogen being slightly more pronounced in V840\,Oph. 
Accordingly, fluorescence via the Bowen mechanism can probably be discarded
as major excitation mechanism, whereas an enhancement of the CN abundance seems
more likely to explain the strength of this feature.

Although high, the ratio of C\,IV\,$\lambda\lambda$580/1 to the nearby
He\,II\,$\lambda$541, which is 1.5 for V840\,Oph, is lower than 
the 3.5
measured for QU\,Car (Drew et al. \cite{drew+03}) and V840\,Oph might hence 
be more similar to V\,Sge where a ratio of 1.3 has been found by 
Herbig et al.  (\cite{herb+65}) although values of 0.4 have also been reported
(Williams \cite{will83}). 
However, for V840\,Oph, this ratio is certainly affected by the extremely 
strong He\,II\,$\lambda$541 line which in V840\,Oph is only a 
factor three fainter than He\,II\,$\lambda$468.
Comparing instead the ratio 
C\,IV\,$\lambda\lambda$580/1 to He\,II\,$\lambda$468 we get 0.5 for
V840\,Oph which is slightly higher than for QU\,Car. 

To better understand the abundances of the different ions and their
excitation mechanism, the study of UV--spectra will be necessary. Additional
time--resolved spectroscopy of the C\,IV  and C\,III lines could help to
clarify the location of the carbon emission source within the system.

\begin{table*}
\caption{\label{eqwtab} Equivalent widths, 
and FWHM of all identified emission
lines in the 2002 and 2003 spectrum of V840\,Oph. For 2003 also the line 
flux as measured and dereddened (same units) is given. 
Note that the uncertainty of the line flux describes the 
uncertainty of the relative flux in the line and does not include the
photometric error.}
\begin{tabular}{l c c c c c c c}
\hline
\hline
 \noalign{\smallskip}
           & \multicolumn{2}{c}{April 2002} & ~ ~ ~ ~ & \multicolumn{4}{c}{February 2003}\\
Transition & FWHM [nm] & $-W$ [nm]  & & FWHM [nm] & $-W$ [nm]  & F [$10^{-18}$\,W\,m$^{-2}$] & dereddened \\
 \noalign{\smallskip}
\hline
 \noalign{\smallskip}
 H$_\alpha$                 & $2.44 \pm 0.03$ & $1.55 \pm 0.15$ & & $3.18 \pm 0.03$ & $1.66 \pm 0.05$ & $0.92 \pm 0.04$ & 2.28 $\pm$ 0.10 \\
 H$_\beta$                  & $2.15 \pm 0.08$ & $1.35 \pm 0.12$ & & $2.45 \pm 0.05$ & $1.51 \pm 0.06$ & $1.37 \pm 0.04$ & 5.22 $\pm$ 0.15 \\
 H$_\gamma$                 &                 &                 & & $2.63 \pm 0.08$ & $1.58 \pm 0.06$ & $1.84 \pm 0.04$ & 8.30 $\pm$ 0.18 \\
 He\,I$\lambda$668          &                 & $< 0.1$         & & $3.20 \pm 0.50$ & $0.27 \pm 0.10$ & $0.15 \pm 0.10$ & 0.36 $\pm$ 0.24 \\
He\,I$\lambda$588 $^{*)}$   &                 & $< 0.1$         & & $ > 1.50 $ & $ > 0.12      $ & $> 0.08$ & \\
He\,I$\lambda$447           &                 &                 & & $2.47 \pm 0.08$ & $0.38 \pm 0.05$ & $0.42 \pm 0.03$ & 1.82 $\pm$ 0.13 \\
 He\,II $\lambda$541        & $1.74 \pm 0.07$ & $0.4 \pm 0.1$   & & $2.55 \pm 0.04$ & $0.61 \pm 0.05$ & $0.45 \pm 0.03$ & 1.43 $\pm$ 0.09 \\
He\,II $\lambda$469 $^{**)}$&                 &                 & & $2.16 \pm 0.15$ & $1.50 \pm 0.07$ & $1.45 \pm 0.05$ & 5.84 $\pm$ 0.20 \\
He\,II $\lambda$454         &                 &                 & & $2.47 \pm 0.08$ & $0.37 \pm 0.05$ & $0.40 \pm 0.04$ & 1.69 $\pm$ 0.17 \\
C\,III $\lambda$570         &                 & $< 0.1$         & & $2.57 \pm 0.30$ & $0.30 \pm 0.04$ & $0.21 \pm 0.02$ & 0.62 $\pm$ 0.06 \\
 C\,IV $\lambda\lambda$580/1& $1.71 \pm 0.02$ & $0.7 \pm 0.1$   & & $2.46 \pm 0.04$ & $0.96 \pm 0.03$ & $0.65 \pm 0.02$ & 1.89 $\pm$ 0.06 \\
 Bowen blend $^{**)}$       &                 &                 & & $3.21 \pm 0.15$ & $1.64 \pm 0.07$ & $1.64 \pm 0.05$ & 6.70 $\pm$ 0.20 \\
 \noalign{\smallskip}
\hline
\end{tabular}
\\
$^{*)}$ {\tiny This He\,I line is disrupted by the nearby Na absorption line.}
\\
$^{**)}$ {\tiny The properties have been derived by fitting five ion components to 
these blended lines (see text and Fig.~\ref{bowen_fit}).}
\end{table*}

So far, one can certainly
conclude that the strength of the Bowen blend as well as the strength
of He\,II hints towards
a higher than solar metallicity in the accreted matter. Therefore 
the secondary in V840\,Oph must be an evolved star.
Hence, the overabundance of carbon, similar to QU\,Car, can be similarly 
explained
by assuming a secondary carbon star. 

Further evidence for a carbon star secondary might come from signatures 
of the secondary in the spectrum of V840\,Oph. 
To search for these features, we have fitted a 
$\lambda ^{\alpha}$ power law to the blue part of the spectrum, leaving out 
the line features. For the dereddened spectrum, we derive $\alpha = 3.303(3)$.
After subtracting this slope from the spectrum we derive the flat spectrum 
plotted in Fig.~\ref{slope_plot}. 
Only a slight enhancement is found at the red end
that could be due to the red continuum of the secondary but might also be
and artificial fit residual.
Three absorption features (Na and two unidentified bands as marked in  
Fig.~\ref{slope_plot})
are found that are also present in the
2002 data and that we hence concluded to be real, although the two bands
are just visible above the noise and we cannot derive their strength. 
Furthermore, the Na line might partly be interstellar. 
However, if all the Na was interstellar, 
its observed equivalent width of 0.19(4)\,nm, which can be considered as a 
lower limit since the line is partly filled up by the nearby 
He\,I\,587.6 emission, would yield an extinction of 
$E_{B-V} > 1.7$ (Zwitter \& Munari \cite{zwit+98}) which is certainly too high.
To match the photometrically derived reddening of $E_{B-V} = 0.4$, at least 
half of the equivalent width has to be due to stellar absorption.

In an attempt to match the absorption features
we find that late type K--stars as well as M--stars show
deep TiO bands with respect to the features present in our data which then
should be visible in the red part of the spectrum. Very early type K--stars as
well as G--stars do not have such a pronounced Na (589\,nm) doublet as to
account for what is observed in the spectrum of V840\,Oph. 
Although spectroscopic observation in the NIR are needed to
improve the classification, we tentatively conclude that the secondary
is best represented by a mid--type K--star.
This is comparable to what has been found for QU\,Car, where 
a late R--star has been suggested as secondary (Drew et al. \cite{drew+03}).

\subsection{Other results from the spectral analysis}

In Table \ref{eqwtab}, the equivalent width and FWHM in both data sets are listed 
for all identified emission lines. For the 2003 data also the
line fluxes as observed and dereddened with $E_{\rm B-V} = 0.4$ 
are given. With values around 2.0\,nm, the FWHM is quite narrow
for a cataclysmic variable.
This might indicate that the system is seen at a low inclination.

A rough comparison of the features already reveals an interesting finding:
The equivalent widths of all lines common in both spectra are systematically
higher
in the 2003 data than in 2002. Furthermore, although faint, He\,I is present 
in this spectrum at $\lambda = $588\,nm and $\lambda =$668\,nm which were not 
detected in 2002.
Both these facts indicate that either V840\,Oph was in a
higher accretion state 
during the 2002 observation compared to 2003, or there is
a dependence of the spectral features on the orbital phase. With the present
data we can not decide between these possibilities.

In order to derive information on the possible temperature range of V840\,Oph, 
we have measured the Balmer decrement, which
is defined as ratio of line intensities $\rm H_\alpha : H_\beta : H_\gamma$. 
We have computed the ratios of line fluxes of $\rm H_\alpha$ and $\rm H_\gamma$
over $\rm H_\beta$ for the 2003 data set and derive 
$\rm H_\alpha/H_\beta = 0.67$ (0.45 dereddened)
and $\rm H_\gamma/H_\beta = 1.34$ (1.59). Allowing for the fact that 
the reddening might be over--corrected, the uncorrected data still yield a 
limit for the Balmer decrement and an accordingly lower limit for the 
temperature and density. 
This highly inverted Balmer decrement indicates that the
line emission has its origin in a hot and dense environment.
A comparison of the lines as observed (without reddening correction)
with model data from
Williams (\cite{will91}) shows indeed that the derived values do not fit in his grid,
which is computed for temperatures up to 15000\,K, but that
an extrapolation towards higher temperatures above 30000\,K is necessary to match both
ratios and the equivalent widths of the lines. Similar high temperatures are
necessary to explain the strength of He\,II compared to the lack of He\,I. At least
20000\,K are estimated for the line--emitting region to maintain the He ionisation.

The presence of high 
excitation lines and especially the high ratio HeII$\lambda$468/H$\beta \approx 1.0$, the absence of He\,I in the 2002 data and its
weakness in 2003, and 
the visibility of absorption lines of the secondary, 
suggest V840\,Oph to be a magnetic system with no or only a very weak disc.
This is supported by the inverted Balmer decrement which also points towards
a magnetic classification (see eg. Schachter et al. \cite{scha+91};
Ferrario \& Wehrse, \cite{ferr+99}).
However, the lack of any pronounced cyclotron lines in the observed part of the
spectrum indicates that
the magnetic field can only be of moderate strength.

\section{Summary}
Using multi colour diagrams for a first selection and optical spectroscopy
for the final confirmation we could identify the classical nova V840\,Oph with
an object of V=19.$^{\rm m}$32 at the position 
$\alpha = 16^{\rm h}54^{\rm m} 43.^{\rm s}9$,
$\delta = -29^{\circ} 37^{\prime} 27^{\prime\prime}$ (J2000). 

The spectrum of V840\,Oph contains an unusually strong C\,IV emission 
at $\lambda = 580/1$\,nm. Furthermore the Bowen blend is strong and 
carbon--dominated. We have concluded that similar to QU\,Car, the spectral
features are best explained by assuming a carbon star as secondary.

V840\,Oph is probably a low inclination system. This highly inverted Balmer 
decrement
and the presence of high excitation lines suggest the system to have extremely
high temperatures above 30000K. 
We furthermore find spectroscopic similarities to magnetic old novae,
suggesting a corresponding scenario for V840\,Oph.
More detailed investigations of this system  
are desirable to confirm its nature.

\acknowledgement{We would like to thank George Hau and Michael Sterzik, who 
kindly performed the spectroscopic observations of V840 Oph in 2002, 
Robert Schwarz for information on magnetic CVs, and the referee Janet Drew for 
helpful comments. We acknowledge that this research has 
made use of the Simbad database operated at CDS, Strasbourg, France.
}


\begin{thebibliography}{}
\bibitem[1920]{bail20}
Bailey, S.I. 1920, AN 210, 375 
\bibitem[1975]{mccl+75}
McClintock, J.E., Canizares, C.R., \& Tarter, C.B. 1975, ApJ 198, 641
\bibitem[2003]{drew+03}
Drew, J.E., Hartley, L.E., Long, K.S., \& van der Walt, J. 2003, MNRAS 338, 401
\bibitem[1981]{duer81}
Duerbeck, H.W. 1981, PASP 93, 165
\bibitem[1987]{duer87}
Duerbeck, H.W. 1987, Space Sci. Rev., 45, 1
\bibitem[1999]{ferr+99}
Ferrario,L. \& Wehrse, R. 1999, MNRAS 310, 189
\bibitem[1982]{gill+82}
Gilliland, R.L. \& Phillips, M.M. 1982, ApJ 261, 617
\bibitem[1994]{harr+94}
Harrison, T.E. \& Gehrz, R.D. 1994, AJ 108, 1899
\bibitem[1965]{herb+65}
Herbig, G.H., Preston, G.W., Smak, J., \& Paczynski, B. 1965, ApJ 141, 617
\bibitem[2002]{hoar+02}
Hoard, D.W., Wachter, S., Clark, L.L., \& Bowers, T.P. 2002, ApJ 565, 511
\bibitem[1986]{horn86}
Horne, K. 1986, PASP 98, 609
\bibitem[1995]{king+95}
Kingsburgh, R.L., Barlow, M.J., \& Storey, P.J. 1995, A\&A 295, 75
\bibitem[1992]{land92}
Landolt,  A.U. 1992, AJ, 104, 340
\bibitem[1998]{muna+98}
Munari, U. \& Zwitter, T. 1998, A\&AS 128, 277
\bibitem[1996]{ring+96}
Ringwald, F.A., Naylor, T., \& Mukai, K. 1996, MNRAS 281, 192
\bibitem[1991]{scha+91}
Schachter, J., Filippenko, A.V., Kahn, S.M., \& Paerels, F.B.S. 1991, 
ApJ 373, 633
\bibitem[1921]{shap21}
Shapley, H. 1921, PASP 33, 189 
\bibitem[1995]{warn95}
Warner, B. 1995, Cataclysmic Variable Stars, Camebridge University Press 
\bibitem[1983]{will83}
Williams, G. 1983, APJ Suppl. Ser. 53, 523 
\bibitem[1991]{will91}
Williams, G.A. 1991, AJ 101, 1929
\bibitem[1983]{will+83}
Williams, R.E. \& Ferguson, D.H. 1983, in: {\it Cataclysmic Variables and 
Related Objects}, eds. Livio \& Shaviv, IAU Coll. 72, p.114 
\bibitem[1998]{zwit+98}
Zwitter, T. \& Munari, U. 1998, in: {\it Wild Stars in the Old West}, eds. 
Howell S., Kuulkers E., Woodward C., ASP Conf.Ser. 137, p.35
\end{thebibliography}
\end{document}